\documentstyle[prd,aps]{revtex}

\begin{document}

\draft
\title{Dimensional renormalization: ladders to rainbows}
\author{R.~Delbourgo\cite{Author1}}
\address{Physics Department, University of Tasmania, GPO Box 252C, Hobart,\\
Australia 7001}
\author{A.C.~Kalloniatis\cite{Author2}}
\address{Institut fur Theoretische Physik, University of Erlangen-Nuremberg, \\
Staudtstrasse 7, Erlangen, D-91058 Germany}
\author{G.~Thompson\cite{Author3}}
\address{International Centre for Theoretical Physics, Grignano-Miramare, \\
Trieste, 34100 Italy}
\date{\today }
\maketitle

\begin{abstract}
Renormalization factors are most easily extracted by going to the massless
limit of the quantum field theory and retaining only a single momentum
scale. We derive the factors and renormalized Green functions to {\em all}
orders in perturbation theory for rainbow graphs and vertex (or scattering)
diagrams at zero momentum transfer, in the context of dimensional
regularization, and we prove that the correct anomalous dimensions for those
processes emerge in the limit $D \rightarrow 4$.
\end{abstract}

\pacs{11.10.Gh, 11.10.Jj, 11.10.Kk}

\narrowtext

\section{INTRODUCTION}


The connection between knot theory and renormalization theory\cite{DK} is
one of the more exciting developments of field theory in recent years
because it relates apparently different Feynman diagrams through the common
topology of the associated knots. Thus it serves to explain why
transcendental numbers for the renormalization constants $Z$ occur in some
diagrams\cite{DBDK} and not in others, thereby allowing the Feynman graphs
to be grouped into equivalence classes. Sometimes, in gauge theories the $Z$
factors within a particular class may cancel because of the existence of
Ward identities, leaving a non-transcendental result for $Z$; this happens
in electrodynamics of scalar and spinor particles to fourth order in the
quenched limit and in chromodynamics to third order\cite{DBRDDK}.

The class of graphs which correspond to ladders and rainbows are especially
simple in this connection because they possess trivial knot topologies. Thus
one may anticipate that $Z$-factors for them are particularly easy to
evaluate. Kreimer\cite{DK} has provided rules for extracting them within the
framework of dimensional regularization, through the standard expedient of
finding the simple 1/$\epsilon =2/(4-D)$ pole term arising in products of
functions, after removing lower-order pole terms connected with
subdivergences. Thus vertex diagrams bring in function factors of the type 
\[
_j\Delta (\epsilon )\equiv (p^2)^{\epsilon (j+1)}\int\frac{d^{4-2\epsilon}k%
} {(k^2)^{1+j\epsilon }(k+p)^2}, 
\]
while rainbow graphs lead to products of 
\[
_j\Omega (\epsilon )\equiv (p^2)^{3+j\epsilon-D/2} \int \frac{d^Dk}{%
(k^2)^{2+j\epsilon}(k+p)^2} . 
\]
It has to be said that, although the procedure is straightforward,
extracting the $1/\epsilon $ term in $n$'th order requires considerable
graft. Kreimer has proven that the simple pole in $\epsilon$ is free of
Riemann zeta-functions.

In this paper we shall show that the problem can be solved to {\em all}
orders in perturbation theory for ladders and rainbows\cite{SE}, in the
context of renormalization in dimensional regularization because of two
fortuitous circumstances: (i) the Green function satisfies a differential
equation and (ii) this equation is actually soluble in terms of Bessel
functions. The limit as $D\rightarrow 4$ may then be taken at the end and,
as a useful check, the anomalous dimension properly emerges. (It is a rather
delicate limit, requiring a saddle-point analysis of the integral
representation of the Bessel function, since it looks quite singular.) We
have successfully carried out this program for meson-fermion theories, both
for vertex functions and rainbow diagrams; however we have not succeeded in
solving the problem near $D=6$ for $\phi^3$ theory because the differential
equation is of fourth order and cannot be expressed in terms of standard
functions; nevertheless we can obtain the answer in the limit $x\rightarrow 0
$ or $p\rightarrow \infty $ for $D=6$.

In the next section we treat the vertex diagrams for scalar mesons, while
the following section contains the analysis of the rainbow diagrams. The
appendix contains details of the vector meson case, which are rather more
complicated.

\section{Vertex diagrams}


We shall consider a theory of massless fermions $\psi $ and mesons $\phi$ in 
$D$-dimensions since the purpose of our work is to investigate the behaviour
of the Green function as $D$ tends to 4. Let $\gamma _{[r]}$ signify the
product of $r\,\,\gamma $-matrices (of size $2^{D/2}\times 2^{D/2}$)
normalized to unity, namely $\gamma_{[\mu_1\mu_2\ldots\mu_r]}$
so that we can write the meson-fermion interaction in the form 
\[
{\cal L}_{{\rm int}}=g\bar{\psi}\gamma _{[r]}\psi \phi ^{[r]}, 
\]
where $\phi^{[r]}$ is the corresponding tensor meson field. The equation for
the renormalized tensor vertex function $\Gamma _{[s]}$ at zero meson
momentum, taking out the factor $g$, is 
\begin{equation}
\Gamma _{[s]}(p)=Z\gamma _{[r]}\delta _s^r-ig^2\int \bar{d}^D\!\!q\,\gamma
_{[r]}\frac 1{\gamma .q}\Gamma _{[s]}(q)\frac 1{\gamma .q}\gamma
_{[r^{\prime }]}\Delta ^{rr^{\prime }}(p-q).
\end{equation}
We shall assume that the massless meson propagator above, $\Delta
^{rr^{\prime }}(p-q)$, can be chosen in a Fermi-Feynman gauge so 
\[
\Delta ^{rr^{\prime }}=(-1)^r\eta ^{rr^{\prime }}/(p-q)^2, 
\]
where $\eta $ stands for the diagonal Minkowskian metric pertaining to the
tensor structure, specifically $\eta^{[\mu_1\nu_1}\cdots
\eta^{\mu_r]\nu_{r\prime}}$.

To make further progress we utilize the non-amputated Green function, 
\[
G_{[r]}(p)=\frac 1{\gamma .p}\Gamma _{[r]}(p)\frac 1{\gamma .p}, 
\]
to remain with the `simpler' linear integral equation, 
\begin{equation}
\gamma .pG_{[s]}(p)\gamma .p=Z\delta _s^r\gamma _{[r]}+i(-)^rc_r^sg^2\int 
\bar{d}^D\!\!q\,\,\frac{G_{[s]}(q)}{(p-q)^2},
\end{equation}
where $\bar{d}^D\!\!q \equiv d^D\!\!q/(2 \pi)^D$. The nature of the
couplings in massless theories means that the Green function always stays
proportional to $\gamma _{[s]}$ and can be decomposed into just two pieces%
\cite{f1}, 
\begin{equation}
G_{[s]}(p)=\gamma _{[s]}A(p^2)+\gamma .p\gamma _{[s]}\gamma .pB(p^2).
\end{equation}
On the right-hand side of (2), $c_r^s\gamma_{[s]}=\gamma_{[r]}\gamma_{[s]}
\gamma ^{[r]}$, is essentially an element of the Fierz transformation matrix
for any $D$, given by\cite{RDVP}

\[
c_r^s=(-1)^{rs}\sum_q (-1)^q 
{D-r \choose s-q}
{r \choose q}
. 
\]

We shall convert the integral equation (2) into a differential equation by
taking the Fourier transform. (In fact we could almost have done this from
the word go by writing the equation for the full Green function in
coordinate space.) This manoeuvre produces 
\begin{equation}
\gamma .\partial G_{[s]}(x)\gamma .\overleftarrow{\partial }=Z\delta
_s^r\gamma _{[r]}\delta ^D(x)+ic_r^s(-)^rg^2\Delta _c(x)G_{[s]}(x),
\end{equation}
where the massless meson propagator is $i\Delta _c(x)=\Gamma
(D/2-1)(-x^2+i\epsilon )^{1-D/2}/4\pi ^{D/2}.$ Because the coupling constant
is a dimensionful quantity, we can define a dimensionless strength `$a=$
fine structure constant $/4\pi $', via 
\[
(-)^rc_r^sg^2\Gamma (D/2-1)\equiv 16\pi ^{D/2}\mu ^{4-D}a 
\]
upon introducing a mass scale $\mu $. This simplifies the resulting
expressions, as we can see in the purely scalar case, where there is but a
single term and equation: 
\begin{equation}
\left[ \partial ^2-\frac{4a}{x^2}(-\mu ^2x^2)^{2-D/2}\right] G(x)=-Z\delta
^D(x).
\end{equation}
The equation is readily solved in dimension $D=4$ yielding $G\propto
(-x^2+i\epsilon )^{(-1-\sqrt{1+4a})/2}$. For $D\neq 4$ we can make progress
by passing to a Euclidean metric ($r^2\equiv -x^2),$%
\[
\left[ \frac{d^2}{dr^2}+\frac{D-1}r\frac d{dr}-\frac{4a}{r^2}(\mu
r)^{4-D}\right] G(r)=Z\delta ^D(r). 
\]
The whole point of the manipulation is that one is fortunately able to solve
this equation (for $r\neq 0$ at first) in terms of known functions, namely
Bessel functions. The correct choice of solution, up to an overall factor,
is 
\[
G(r)\propto r^{\epsilon -1}J_{1-1/\epsilon }(\sqrt{-4a}(\mu r)^\epsilon
/\epsilon );\qquad D\equiv 4-2\epsilon . 
\]
because in the limit as $a\rightarrow 0$ we recover the free-field solution $%
r^{2-D}$. For dimensional reasons, let us carry out our renormalization so
that $G(1/\mu)=\mu^{2-2\epsilon}.$ With that convention, the vertex function
reduces to 
\begin{equation}
G(r)=(r/\mu)^{\epsilon -1}J_{1-1/\epsilon }(\sqrt{-4a}(\mu r)^\epsilon
/\epsilon )/J_{1-1/\epsilon }(\sqrt{-4a}/\epsilon ).
\end{equation}
Furthermore the correct singularity for the time-ordered function $\delta
(x) $ emerges if we reinterpret $r^2=-x^2+i\epsilon $ above. We shall not
worry at this stage whether $a$ is positive or negative---the sign can vary
with the model anyway---since we can easily continue the function from $J$
to $I$ as needed.

The problem presents itself: how does the four-dimensional result, with its
anomalous scale $\gamma =\sqrt{1+4a}-1$, emerge from (6) as $\epsilon
\rightarrow 0^{-}$, say? This is clearly a delicate limit because both the
index and the argument of the Bessel function become infinitely large.
Before answering this question, let us note that in a perturbative expansion
of (4), {\em viz}. a small argument expansion of $J$, 
\[
G(r)=r^{2\epsilon -2}\frac{[1+\frac{a(\mu r)^{2\epsilon }}{\epsilon
(2\epsilon -1)}+\frac{a^2(\mu r)^{4\epsilon }}{2\epsilon ^2(2\epsilon
-1)(3\epsilon -1)}+\cdots ]}{[1+\frac a{\epsilon (2\epsilon -1)}+\frac{a^2}{%
2\epsilon ^2(2\epsilon -1)(3\epsilon -1)}+\cdots ]}\text{ }
\]
the poles in $\epsilon $ cancel out to any particular order in $a$. For
instance up to order $a^2$ we obtain as $\epsilon \rightarrow 0$, 
\[
G(r)=r^{-2}[1+2(-a+a^2)\ln \mu r+2a^2(\ln \mu r)^2+\cdots ]
\]
which agrees precisely with the expansion of the anomalous dimension in the
logarithmic terms. Returning to the limit of small $\epsilon $, we will make
use of the saddle point method of obtaining asymptotic expansions of
integrals. Suppose that $f(t)$ has a minimum at $t=\tau $ in the integral
representation, 
\[
F=\frac{-i}{2\pi }\int_C\exp f(t)\;dt.
\]
Then the saddle point method gives 
\[
F=\exp f(\tau )\frac 1{\sqrt{2\pi f^{\prime \prime }(\tau )}}\left[ 1+\frac{%
f^{\prime \prime \prime \prime }(\tau )}{8(f^{\prime \prime }(\tau ))^2}-%
\frac{5(f^{\prime \prime \prime }(\tau ))^2}{24(f^{\prime \prime }(\tau ))^3}%
+\frac{35(f^{\prime \prime \prime \prime }(\tau ))^2}{384(f^{\prime \prime
}(\tau ))^4}+\cdots \right] .
\]
As confirmation of the correctness of the terms above we can verify that the
Debye expansion\cite{AS} of the Bessel function is properly reproduced, 
\begin{eqnarray*}
J_\nu (\nu /\cosh \tau ) &=&\frac{-i}{2\pi }\int_{-i\pi +\infty }^{i\pi
+\infty }dt\,{\rm e}^{\nu (\sinh t/\cosh \tau -t)} \\
&=&\frac{{\rm e}^{\nu (\tanh \tau -\tau })}{\sqrt{2\pi \nu \tanh \tau }}%
\left[ 1+\frac{\coth \tau }{8\nu }\left( 1-\frac 53(\coth \tau )^2\right)
+\cdots \right] ,
\end{eqnarray*}
because the integrand minimum occurs at $t=\tau $. Our case (6) is a variant
of this. Working only to first order in $\epsilon $, and taking $a$ negative
initially, the integrand exponent is 
\[
(\sqrt{-4a}\sinh t+t)/\epsilon +\sqrt{-4a}\ln \mu r\sinh t-t
\]
and is stationary at the complex value $t=\tau $, where 
\[
\cosh \tau =\frac{\epsilon -1}{\sqrt{-4a}(1+\epsilon \ln \mu r)}.
\]
Following through the mathematical steps, and omitting straightforward
details, to order $\epsilon $ we end up with, 
\begin{equation}
G(r)=r^{2\epsilon -2}(\mu r)^{1-\epsilon -\sqrt{1+4a}}\left[ 1-\frac{%
2a\epsilon }{1+4a}\ln (\mu r)+\cdots \right] .
\end{equation}
It is satisfying that this produces the all-orders (in coupling, $a$) result
at 4-dimensions when $\epsilon \rightarrow 0$, with the correct anomalous
scale.

The problem can be treated in much the same way for a pseudoscalar meson
field. The only possible difference is a change in sign of $a$, because of `$%
\gamma _5$' matrix anticommutation. As for the vector case ($r=s=1$), the
procedure produces a pair of coupled equations for the two scalar components 
$A$ and $B$ of the Green function, $G_\mu (p)=\gamma _\mu A(p^2)+ \gamma
.p\gamma _\mu \gamma .pB(p^2)$. A discussion of this case is given in the
appendix, where it is shown that the only easy limit is $D=4$; one finds
after Euclidean rotation that 
\[
A=ar^\beta ,\quad B=br^{\beta +2};\qquad a/b=c(\beta +2)/(\beta -2), 
\]
where 
\[
\beta =-1+\sqrt{5+\sqrt{16+4c+c^2}}\quad {\rm and}\quad c=g^2/2\pi ^2, 
\]
in four dimensions; we have chosen the root which reduces to the free field
solution when $g=0$ although one can contemplate strictly non-perturbative
solutions\cite{AF}.

The difficulty is symptomatic of what happens in $\phi ^3$ theory near six
dimensions; in that case the Green function, $G(p)=\Gamma (p)/p^4$ obeys the
Fourier transformed equation\cite{f12}, 
\begin{equation}
\left[ \partial ^4-4a\frac{(\mu r)^{6-D}}{r^4}\right] G(x)=Z\delta ^D(r).
\end{equation}
This is a differential equation of fourth order in $r$ and its solution
cannot readily be expressed in terms of familiar transcendental functions.
However in the limit as $D\rightarrow 6$, it is quite simple to find the
(power law) solution: 
\[
G(r)\propto r^\beta ;\qquad \beta =-1-\sqrt{5+2\sqrt{4+a}}, 
\]
which correctly reduces to the free-field solution $\beta =-4$ when the
coupling vanishes.

\section{Rainbow diagrams}

We wish to treat the corrections to the fermion propagator in a similar
manner, by considering the rainbow corrections. In such an approximation the
rainbow graphs give rise to a self-energy, which is self-consistently
determined according to 
\[
\Sigma_R(p)=-ig^2\int\frac{\bar{d}^D\!q}{(p-q)^2}\left[ \frac{1}{\gamma.q}- 
\frac{1}{\gamma.q}\Sigma_R(q)\frac{1}{\gamma.q} \right], 
\]
with the unrenormalized propagator determined by 
\[
S_R(p)=\frac{1}{\gamma.p}-\frac{1}{\gamma.p}\Sigma_R(p)\frac{1}{\gamma.p} 
\]
at this level\cite{f2}. This leads to the renormalized rainbow corrected
propagator equation, 
\begin{equation}
\gamma.p S_R(p)\gamma.p=Z_\psi\gamma.p+ig^2\int\frac{\bar{d}^D\!q}{(p-q)^2}
S(q).
\end{equation}
The nature of the massless problem is that one can always write $S_R(p) =
\gamma.p \sigma(p)$, and by Fourier transformation, convert (9) from an
integral equation to a differential equation, 
\[
-i\gamma.\partial\partial^2\sigma(x) = i\gamma.\partial Z_\psi\delta^D(x) +
ig^2\Delta_c(x)\,i\gamma.\partial\sigma(x) 
\]
or 
\begin{equation}
\left[\partial^2+ig^2\Delta_c(x)\right]\partial_\mu\sigma(x) = - Z_\psi
\partial_\mu \delta^D(x)
\end{equation}
Now for any function $f(\sqrt{x^2})$, using the two lemmas, 
\begin{mathletters}
\begin{equation}
\partial_\mu f = x_\mu f^{\prime}/\sqrt{x^2},
\end{equation}
\begin{equation}
\partial_\mu\partial_\nu f = \left(\eta_{\mu\nu}-\frac{x_\mu x_\nu}{x^2}
\right) \frac{f^{\prime}}{\sqrt{x^2}} + \frac{x_\mu x_\nu}{x^2}
f^{\prime\prime},
\end{equation}
we can carry out an Euclidean rotation in order to arrive at the
differential equation for the scalar function ${\cal S} \equiv d\sigma/dr$: 
\end{mathletters}
\begin{equation}
\left[\frac{d}{dr}\left(\frac{D-1}{r}+\frac{d}{dr}\right) + \frac{4a(\mu
r)^{4-D}}{r^2} \right]{\cal S}(r) = -Z_\psi\delta^D(r).
\end{equation}
In 4-D this has the simple solution ${\cal S}(r)\propto r^{-1-2\sqrt{1+a}}$,
in turn implying $S(p) \propto \gamma.p\,p^{-4+2\sqrt{1+a}}$. However it is
in fact possible to solve (12) for any $D$. The proper solution, normalized
to ${\cal S}(1/\mu) = \mu^{3-2\epsilon}$ is 
\begin{equation}
{\cal S}(r) = r (r/\mu)^{\epsilon-2}\frac{J_{1-2/\epsilon} (\sqrt{-4a}(\mu
r)^\epsilon/\epsilon)}{J_{1-2/\epsilon} (\sqrt{-4a}/\epsilon)}.
\end{equation}
To get the rainbow propagator, we must first integrate, $\sigma(r) = \int^r%
{\cal S}(r)\,dr \propto \sum_{m=0}^\infty J_{2-2\epsilon+2m} (\sqrt{-4a}(\mu
r)^\epsilon/\epsilon)$, and then Fourier transform to obtain $%
S_R(p)=\gamma.p\sigma(p)$.

If the mesons are neither scalar nor pseudoscalar, but tensor, the coupling
constant is multiplied by the factor $c_1^r$; that is all.

\section{Conclusions}


We have demonstrated that it is possible to work out the all-orders solution
of Green functions for ladder and rainbow diagrams for any dimension $D$ and
that, in the limit as $D$ approaches the physical dimension, the correct
scaling dimension is obtained. We have exhibited fully how this happens for
scalar theories, but have succeeded only to a limited extent in vector
theories, because the equations are coupled and end up as fourth order ones,
with no transparent expression in terms of standard functions of
mathematical physics. In any event, it is clear from the form of the Green
function that there are no transcendental constants in sight, even when we
expand the answers perturbatively in terms of $\ln(\mu r)$, so that the
renormalization constants are free of them. This confirms the finding of
Kreimer for arbitrary ladder/rainbow order \cite{DBDK} and does not come as
a surprise.

One can extend the ideas here to scattering processes which contain a single
momentum scale, such as fermion-fermion scattering (again ladder graphs) for
any $D$. It is a simple matter of taking the Fourier transform in particular
channels and converting the momentum integral equations to differential ones
in coordinate space. We shall not labour the issue in this paper since the
steps are fairly obvious and can easily be filled in by the reader. What we
have not solved for any $D$ is the case of crossed ladders, when the kernel
will presumably lead to transcendental $Z$ constants; that is a task for the
future.

\acknowledgments
We would like to thank the Australian Research Council for providing
financial support in the form of a small grant during 1995---when the
majority of this work was carried out.

\appendix

\section*{The vector case}

The vector vertex function 
\[
G_\mu (p)=A(p^2)\gamma _\mu +\gamma .p\gamma _\mu \gamma .p\,B(p^2), 
\]
upon Fourier transformation and tracing with $\gamma _\nu $, produces the
coordinate space equation, 
\[
(\partial ^2\eta _{\mu \nu }-2\partial _\mu \partial _\nu )A+\partial ^4\eta
_{\mu \nu }B=Z\eta _{\mu \nu }\delta (x)+ig^2(D-2)\Delta _c(x)[\eta _{\mu
\nu }A+(\partial ^2\eta _{\mu \nu }-2\partial _\mu \partial _\nu )B]. 
\]
Using lemmas (11), and identifying the terms multiplying $\eta _{\mu \nu }$
and $x_\mu x_\nu ,$ we arrive at the pair of coupled equations, 
\[
{\cal O}\left[ -A+{\cal O}B\right] +\frac 22\frac{dA}{dr}=ig^2(D-2)\Delta
_c\left[ A+\frac 2r\frac{dB}{dr}-{\cal O}B\right] , 
\]
\[
{\cal Q}A=ig^2(D-2)\Delta _c\,{\cal Q}B, 
\]
where ${\cal O}\equiv (\frac{d^2}{dr^2}+\frac{D-1}r\frac d{dr})$ and ${\cal Q%
}\equiv (\frac{d^2}{dr^2}-\frac 1r\frac d{dr})$. We have not suceeded in
solving these equations in terms of familiar functions for $D\neq 4$.
However in 4-D, one can make considerable progress by looking for a
power-law solution of the type, $A(r)=ar^\beta $, $B(r)=br^{\beta +2}$.
Simple calculation reveals that a solution exists provided that 
\[
a/b=c(\beta +2)/(\beta -2),\quad c^2+4c-\beta (\beta -2)(\beta +2)(\beta
+4)=0,\qquad c\equiv g^2/2\pi ^2. 
\]
The quartic in the power exponent $\beta $ is fortunately simple to solve in
terms of the coupling (or $c$), the answer being 
\[
\beta =-1+\sqrt{5+\sqrt{16+4c+c^2}}=2+c/3-c^2/54\cdots , 
\]
so that $a/b\simeq 12+5c/3+\cdots $.


\begin{references}
\bibitem{f1}  Equation (3) is merely the most convenient way of writing out
the decomposition; it can be transformed into the form $A\gamma _{[\mu
_1\ldots \mu _t]}+C\sum_{j=1}^tp_{\mu _j}p^\lambda \gamma _{[\mu _1..\mu
_{j-1}\lambda \mu _{j+1}..\mu _t]}$, by using the gamma-matrix commutation
relations. Here the tensor structure is more explicit. Note that $C$ or $B$
terms do not have an independent existence when $r=0$ (scalar) or when $r=D$
(pseudoscalar).
\bibitem{f12}  Remember that for an interaction $g\phi ^3$ in $D$
dimensions, $g$ has mass dimension $3-D/2$, while the propagator behaves as $%
1/r^4$. That is why the strength $a$ is dimensionless in equation (8).
\bibitem{f2}  Chains of self-energies are {\em not} being summed here. Were
one to attempt that, it would be necessary to tackle the full
Dyson-Schwinger equation $S^{-1}(p)=\gamma .p+\Sigma (p),$ which would lead
to a non-linear equation for the propagator.
\end{references}
\end{document}